\documentclass[aps,pre,reprint,onecolumn,superscriptaddress,nofootinbib]{revtex4-2}
\usepackage{amsmath, amsfonts, amssymb}
\usepackage{dsfont}
\usepackage{placeins} 
\usepackage{mathtools}
\usepackage{booktabs}
\usepackage{bm}
\usepackage{braket}
\newcommand{\abs}[1]{\left|#1\right|}
\usepackage[makeroom]{cancel}
\usepackage{ragged2e}
\usepackage{graphicx} 
\usepackage[dvipsnames]{xcolor}
\usepackage{orcidlink}
\usepackage{orcidlink}
\hypersetup{
    colorlinks=true,
    citecolor=blue,
    linkcolor=blue,
    urlcolor=blue
}
\begin{document}

\title{The peculiar response of Kelvin-Voigt chains  with a free end}

\author{Rupayan Saha \orcidlink{0009-0006-0082-807X}}
\email{rupayan.saha@uni-goettingen.de}
\affiliation{Institut f{\"u}r Theoretische Physik, Georg-August Universit{\"a}t G{\"o}ttingen, 37077 G{\"o}ttingen, Germany}

\author{Matthias Kr{\"u}ger \orcidlink{0000-0001-5015-935X}}
\email{matthias.kruger@uni-goettingen.de}
\affiliation{Institut f{\"u}r Theoretische Physik, Georg-August Universit{\"a}t G{\"o}ttingen, 37077 G{\"o}ttingen, Germany}

\date{\today}

\begin{abstract}
%\justifying
We exactly solve a model of a heterogeneous chain of overdamped, harmonically coupled particles with momentum-conserving dissipation.  Despite being governed by a non-symmetric drift operator, the system admits an analytical diagonalization by use of a forward-difference transformation. In case of one free end, the response matrix shows a peculiar staircase form: the response of particle $i$ to a force acting on particle $j$ is independent of the properties and the length of the chain-part between $i$ and $j$.  For rank-deficient interaction matrices,   the state space is decomposed  into free and constrained subspaces. We demonstrate that this separation has clear physical consequences: the free subspace governs steady state responses, while the constraint subspace governs the relaxation after cessation of forcing. We further show that a Maxwell chain arises as a singular limiting topology of the Kelvin--Voigt chain. These results establish a  framework for analyzing heterogeneous overdamped dynamics with momentum conservation.
\end{abstract}

\maketitle

\section{Introduction}
 
The one-dimensional harmonic chain is a paradigmatic model for collective dynamics in many-body systems, combining analytical tractability with broad physical relevance. Its legacy spans from Huygens' foundational works on synchronized pendulums \cite{Huygens336} and the study of small oscillations in the eighteenth century \cite{goldstein_mechanics} to the early twentieth-century development of lattice dynamics and phonon theory in solids \cite{einstein1907, BornKarman1912, debye1912, kittel2004introduction}. In the latter half of the twentieth century, the  growing interest in soft matter and polymer physics led to the modification of the same framework into the overdamped limit \cite{rouse1953theory}, eliminating inertia in favor of dissipation with respect to a surrounding ``bath" \cite{CALDEIRA1983587}, acting as a \emph{local} momentum sink. In this standard case with homogeneous interaction (stiffness) and dissipation (friction), the normal modes coincide with those of the inertial harmonic chain \cite{doi1988theory} with oscillation frequencies replaced by relaxation rates. 

However, allowing for spatial heterogeneity in elastic or dissipative properties, as, e.g., encountered in chemically heterogeneous polymers, coarse-grained biomolecular models, or spatially heterogeneous hydrodynamic environments,  often precludes a normal mode decomposition, and hence motivates numerical approaches \cite{Rolls2017, Hung2018}, with  few analytical exceptions \cite{chen2023}. Furthermore, incorporating \emph{internal} friction, reflecting dissipation within the polymer or mediated by the solvent \cite{khatri2007, makarov2013, rajarshi2021}, further anticipate the emergence of relative and momentum-conserving dissipation, characteristic of hydrodynamic descriptions \cite{Zimm1956}. 
A qualitatively new regime emerges when momentum conservation is enforced by modeling dissipation as arising \textit{solely} from internal friction within the surrounding medium \cite{Kailasham2026DPD}, rather than from coupling to a fixed background. Such momentum-conserving dissipation, naturally realized in dissipative particle dynamics (DPD) \cite{Hoogerbrugge_1992,Espanol_1995}, and related approaches \cite{PEspanol_1997,BonetAvalos_1997, Huang2007, Das2024}, gives rise to relative friction, whereby the dissipative force on a degree of freedom depends on its relative motion with respect to neighboring degrees of freedom.

Remarkably, although spatial heterogeneity in general impedes analytical treatment, momentum-conserving dissipation with nearest-neighbor structure keeps the system  exactly solvable \cite{Guo2004EigenViscoelastic, SerraAguila2019, Trcala2024}. We show here that this is achieved by introducing a forward-difference transformation that maps the dynamics from site variables (physical displacements) to bond variables (relative displacements). In the present setup, both the stiffness (\(\mathrm{K}\)) and dissipation (\(\Gamma\)) matrices acquire a compatible structure that can be expressed in terms of difference operators, leading to an exact decoupling of the dynamics. Consequently, the drift matrix, $\mathrm{W} = \Gamma^{-1} \mathrm{K}$, remains diagonalizable even in the presence of heterogeneity. Thus, the same physical ingredients that introduce complexity -- spatial variation and hydrodynamic consistency -- also impose sufficient structure to restore analytical tractability.

The same structure manifests in strikingly nonlocal mechanical responses. In particular, we demonstrate that chains with an open end exhibit particularly counter-intuitive responses: the response of particle $i$ to a force acting on particle $j$ does not depend on the properties or length of the chain in between $i$ and $j$, encoded in a staircase pattern in the response matrix. We refer to this as an ``X-ray vision" property, where the response effectively sees through intermediate heterogeneity. 

While the staircase response already reveals the peculiar signal propagation, a deeper physical distinction emerges when we apply this framework to viscoelastic fluids which requires an underconstrained, rank-deficient interaction matrix \(\mathrm{K}\), possessing free modes associated with irreversible rearrangements (reminiscent of a fluid), while retaining constrained modes capable of opposing elastic deformation (resembling a solid) \cite{PHANTHIEN20166359, Tong2023}.  The resulting structure of $\mathrm{W}$ then admits an orthogonal decomposition into free modes enabling fluid-like instantaneous response and constrained modes producing a delayed mechanical response, despite the absence of inertia. In particular, we demonstrate that the free modes exclusively govern the steady state response, while the constrained modes exclusively govern the relaxation after cessation of forcing. This allows for a clean (experimental) distinction between these modes.  We exemplify these findings in a setup of a sheared fluid with viscoelastic components.

The paper is organized as follows. 
Sec.~\ref{pref} is a preface, aimed to induce some intuition for the findings to follow. In Sec.~\ref{model}, we introduce the model, formulate the problem, and derive the formal solution. In Sec.~\ref{eigen}, we cast the dynamics as an eigenvalue problem and obtain an exact diagonalization along with the corresponding eigenmodes. In Sec.~\ref{cor_res}, we discuss the linear response of the system to external perturbations. In Sec.~\ref{sp_case}, we focus on the case of a rank-deficient interaction matrix, which enables an orthogonal decomposition of the state space into constrained and free modes and illustrate this structure through a physically motivated example in Sec.~\ref{example}. In Sec.~\ref{maxwell_limit}, we discuss the Maxwell chain as a singular limiting case. In Sec.~\ref{FDT}, we relate the response to equilibrium fluctuations through the Fluctuation--Dissipation Theorem. Finally, Sec.~\ref{con} presents our conclusions. 

\section{Preface: The massless undamped chain} \label{pref}
\begin{figure*}
    \centering
    \includegraphics[width=0.85\linewidth]{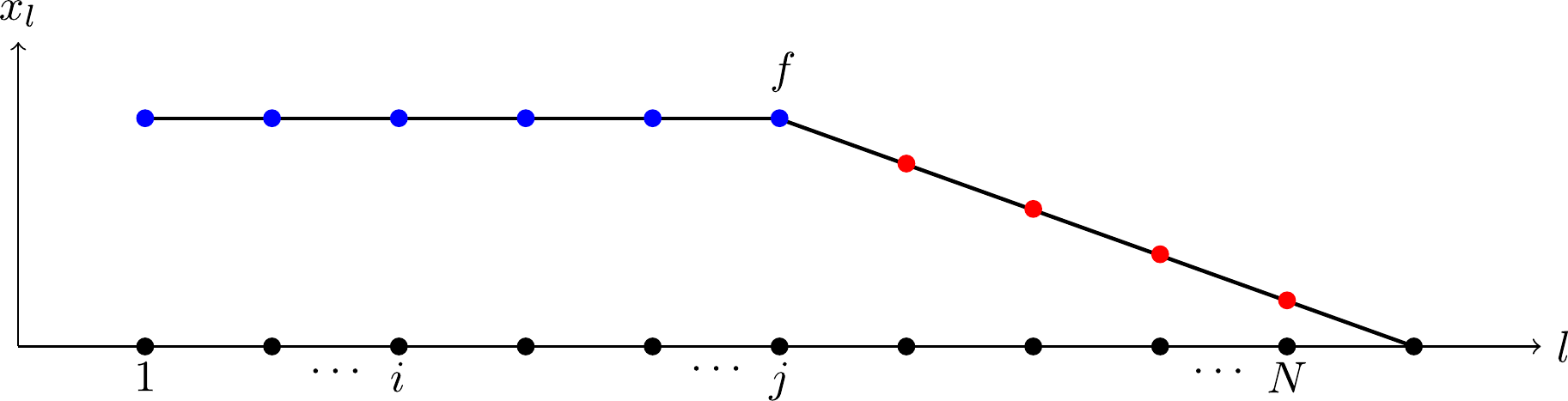}
    \caption{\justifying \textbf{Displacement profile in a massless undamped harmonic chain} of length \(N\), with left end free and the right end (site $N+1$) fixed. Lower dots show the sites in absence of external forces, where $x_l=0$ for all $l$. Upper dots sketch the positions after application of an external force \(f\) acting on site \(j\). This yields $x_l=x_j$ for all \(l < j \), implying a staircase form of the response matrix.}
    \label{fig_preface}
\end{figure*}

Consider a chain of harmonically coupled particles with one end free and one end fixed, and the particles in the chain to be massless and frictionless. In this case, if a force acts on particle $j$, all particles $i$ with $i<j$ will respond \emph{in the same way} as particle $j$, as they are force free if relative distances are kept (see Fig.~\ref{fig_preface}).
In particular, the response of particle $i<j$ is independent of the properties or the length of the part of the chain  in between $i$ and $j$. As the response is reciprocal, this statement can be generalized: The response of an arbitrary particle $i$ to a force acting on an arbitrary particle $j$ is independent of the properties or the length of the part of the chain in between $i$ and $j$ . Inversion of the interaction matrix demonstrates similar properties  for the static equilibrium correlations of the mentioned chain. 

While a massless system may be hard to realize in practice, we  demonstrate in this manuscript that a chain of overdamped Kelvin--Voigt units shows, among other things, the mentioned properties, which has various experimental ways of realization.     

\section{Model} \label{model}

\begin{figure*} 
    \centering
    \includegraphics[width=1\linewidth]{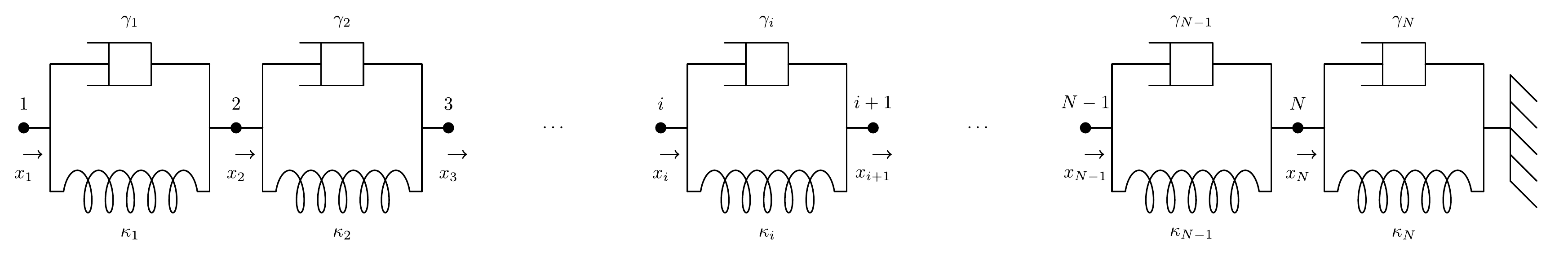}
    \caption{\justifying \textbf{Schematic diagram of the model in Eq.~\eqref{explicit_EOM}},  consisting of $N$ interacting  DoFs with nearest-neighbor interaction \(\kappa_i\), and relative friction \(\gamma_i\). The left end is free while the right end is fixed, $x_{N+1}=0$. The model thus represents a chain of Kelvin--Voigt units \cite{Guo2004EigenViscoelastic, SerraAguila2019, Trcala2024}. }
    \label{fig_model}
\end{figure*}

We consider a one-dimensional chain of $N$ interacting degrees of freedom (DoFs) with nearest-neighbor interactions. Each DoF $x_i(t)$ is coupled elastically with a harmonic spring of stiffness $\kappa_i$, and connected dissipatively with a dashpot of friction coefficient $\gamma_i$ to its neighbor, $x_{i+1}(t)$ (see Fig.~\ref{fig_model}). Additionally, \(x_i\) is subject to  a  force $f_i$.  The equation of motion for degree $x_i$ is given by,
\begin{align} \label{explicit_EOM}
\gamma_i (\dot{x}_{i} - \dot{x}_{i+1}) + \gamma_{i-1} (\dot{x}_{i} - \dot{x}_{i-1}) + \kappa_i (x_{i} - x_{i+1}) + \kappa_{i-1} (x_{i} - x_{i-1}) = f_{i}; \quad \forall i =1,\dots,N.  
\end{align}
We consider a free boundary condition at one end, \( x_0 = x_1 \), and a fixed  boundary condition at the other, \( x_{N+1}=0\). Thus, the system can be regarded as a collection of Kelvin--Voigt units \cite{Guo2004EigenViscoelastic, SerraAguila2019, Trcala2024}.
The set of equations in Eq.~\eqref{explicit_EOM} can be formulated compactly in terms of a state vector, $\mathbf{x} = \{x_1, x_2, \dots,x_N\}^\top \in \mathbb{R}^N $, comprising the $N$ DoFs,   
\begin{equation} \label{vect_EOM}
    \Gamma \dot{\mathbf{x}} + \mathrm{K} \mathbf{x} = \mathbf{f},        
\end{equation}
where, the force \( \mathbf{f} \in \mathbb{R}^N \), is given by, \( \mathbf{f} = \{ f_1, \dots, f_N\}^\top \). The matrices, $\Gamma , \mathrm{K} \in \mathbb{R}^{N \times N} $, encoding \textit{friction} and \textit{stiffness}, respectively, are tridiagonal, reflecting the nearest-neighbor interactions via dashpots and springs, %(cf. Appendix \ref{App_Explicit} for details)
\begin{align} 
\Gamma_{ij} &= (\gamma_i + \gamma_{i-1})\,\delta_{ij}
- \gamma_i \,\delta_{i+1,j}
- \gamma_{i-1} \,\delta_{i-1,j}, \label{friction_explicit} \\
\mathrm{K}_{ij} &= (\kappa_i + \kappa_{i-1})\,\delta_{ij}
- \kappa_i \,\delta_{i+1,j}
- \kappa_{i-1} \,\delta_{i-1,j}. \label{stiffness_explicit}
\end{align}
We consider $\gamma_i > 0$, which makes $\Gamma$ positive definite and ensures  that the inverse, $\Gamma^{-1}$, exists. Moreover, we consider $\kappa_i \geq 0$, i.e., we  allow zero values for $\kappa_i$, to be discussed in detail in Sec.~\ref{sp_case}. The formal solution of Eq.~\eqref{vect_EOM} for time $t$ is given by, 
\begin{align} \label{MOUP_solution}
   \mathbf{x}(t) = e^{-\mathrm{W} (t-t_0)} \mathbf{x}(t_0) + \int_{t_0}^{t} ds e^{-\mathrm{W} (t-s)} \Gamma^{-1}\mathbf{f} (s). 
\end{align}
$\mathrm{W} = \Gamma^{-1} \mathrm{K}$ is  the \textit{drift} matrix, and \( \mathbf{x}(t_0) \), with $t_0<t$, is the initial state. 
Thus, the time evolution is governed by the propagator \( \exp{(-\mathrm{W} t)} \), requiring identification of  the eigenmodes of $\mathrm{W}$.
We turn to the problem of diagonalizing $\mathrm{W}$ in the next section.

\section{Eigenmodes} \label{eigen}

The drift matrix, $\mathrm{W} = \Gamma^{-1}\mathrm{K}$, is non-symmetric, $\mathrm{W}^\top \neq \mathrm{W}$, since $\Gamma$ and \(\mathrm{K}\) do not commute,  $[\Gamma,\mathrm{K}] \neq 0$. Therefore, it is not  \textit{a priori} clear whether its eigenvectors form a well-behaved basis.
To establish such a representation, one requires completeness together with orthogonality, which is determined by the choice of inner product. Equipping the space with the $\Gamma$-weighted inner product, $\mathbf{a}^\top \Gamma  \mathbf{b} \equiv \braket{a| \Gamma| b} := \braket{a|  b}_\Gamma $ \footnote{ In this section, we introduced the Dirac (bra-ket) notation for the ease of analysis: \( \ket{a} \equiv \mathbf{a}, ~  \bra{a} \equiv \mathbf{a}^\top\). }, renders the operator $\mathrm{W}$ self-adjoint, i.e., one may show,
\begin{align} \label{inner_product}
\braket{ \mathrm{W} a| b}_\Gamma  = \braket{ a| \mathrm{W}^\top \Gamma| b} =  \braket{ a| \mathrm{K}| b} = \braket{ a| \Gamma \mathrm{W} |  b} =  \braket{  a| \mathrm{W} b}_\Gamma. 
\end{align}
By virtue of the spectral theorem \cite{axler2024linear} for self-adjoint operators in finite-dimensional vector spaces, it then follows that $\mathrm{W}$ is diagonalizable with real eigenvalues, and its eigenvectors can be chosen to form a complete orthonormal basis with respect to $\braket{\cdot | \cdot }_\Gamma$. Thus, dissipation $\Gamma$ arises naturally as the suitable metric within the vector space \(\mathbb{R}^N \). 

Eq.~\eqref{inner_product} relies only on the positivity of $\Gamma$ and the symmetry of $\mathrm{K}$, so that for any such case, $\mathrm{W}$ has eigenvectors that form a complete orthonormal basis.  

While the spectral theorem guarantees the existence of a complete orthonormal eigenbasis, it does not, as such, provide an explicit construction of the eigenmodes when $\Gamma$ and $\mathrm{K}$ do not commute. In heterogeneous systems, e.g., cases where $\gamma_i$ or $\kappa_i$ vary with $i$, this typically renders the analytical diagonalization of $\mathrm{W}$  intractable. Nonetheless, in the case of Eq.~\eqref{explicit_EOM}, the tridiagonal structure of matrices, $\Gamma, \mathrm{K}$, in Eqs.~\eqref{friction_explicit}, \eqref{stiffness_explicit} enables analytical diagonalization of $\mathrm{W}$. To exploit this structure, we introduce the forward difference (discrete derivative) operator $\mathrm{L}$, given by, 
\begin{align} \label{L_matrix}
    \mathrm{L}_{ij} = \delta_{ij} - \delta_{i,j+1}; \quad (\mathrm{L} \mathbf{x})_i = x_i- x_{i+1}.
 \end{align}
This operator maps site variables  to bond variables (relative displacements), in which the dynamics simplifies considerably. Mathematically, \(\mathrm{L}\) allows to express $\Gamma$ and $ \mathrm{K}$ as congruence transformations \cite{weissteinCongruence}, 
\begin{align}  \label{congruence}
   \Gamma = \mathrm{L}^\top \Lambda_{\Gamma} \mathrm{L} , \quad \mathrm{K} = \mathrm{L}^\top \Lambda_{\mathrm{K}} \mathrm{L},
\end{align}
with, \( \Lambda_{\Gamma} = \operatorname{diag} \{ \gamma_1,\gamma_2,\dots,\gamma_N \} \), and \( \Lambda_{\mathrm{K}} = \operatorname{diag} \{ \kappa_1,\kappa_2,\dots,\kappa_N \} \). Note that Eqs.~\eqref{congruence} are not to be confused with diagonalization of $\Gamma$ and $ \mathrm{K}$, as  $\mathrm{L}^\top\not=\mathrm{L}^{-1}$. $\mathrm{L}$ however allows  to diagonalize $\mathrm{W}$, by use of the representations of $\Gamma$ and $\mathrm{K}$  given in Eq.~\eqref{congruence}, 
\begin{align} \label{similarity_transform}
   \mathrm{W} = \Gamma^{-1}\mathrm{K} = \mathrm{L}^{-1} \Lambda_{\Gamma}^{-1} \Lambda_{\mathrm{K}} \mathrm{L}. %\equiv(\mathrm{L}^{-1})\Lambda \mathrm{S}^{-1},
\end{align}
This is  a similarity transform \cite{arfken2011mathematical} that diagonalizes $\mathrm{W}$ to $\Lambda$,  
\begin{align} \label{eigenspectrum}
   \Lambda :=  \Lambda_{\Gamma}^{-1} \Lambda_{\mathrm{K}} = \operatorname{diag} \{\dfrac{\kappa_1}{\gamma_1},\dfrac{\kappa_2}{\gamma_2}, \dots, \dfrac{\kappa_N}{\gamma_N} \} .
\end{align}
We identify $\mathrm{L}^{-1}$ from Eq.~\eqref{similarity_transform} to consist of the eigenvectors of \(\mathrm{W}\) as columns (see Appendix~\ref{App_Explicit}).
The eigenvectors \( \{ \ket{n} \} \) satisfy,
\begin{align} \label{eigenvector}
   \mathrm{W} \ket{n} = \lambda_n \ket{n} ; \quad \lambda_n := \dfrac{\braket{n| \Gamma \mathrm{W}| n}}{\braket{n|\Gamma| n}} = \dfrac{\braket{n|\mathrm{K}| n}}{\braket{n|\Gamma| n}} = \dfrac{\kappa_n}{\gamma_n}.
\end{align}
The eigenvalues \( \{ \lambda_n  \} \) in Eq.~\eqref{eigenspectrum} correspond to relaxation rates of the associated eigenmodes. 

Moreover, the eigenvectors are orthogonal with respect to the inner product, 
\begin{align} \label{orthogonal}
    \braket{n|\Gamma|m} = \gamma_n \delta_{nm},
\end{align}
and form a complete set of basis,
\begin{align} \label{complete}
    \sum_n \dfrac{1}{\gamma_n} \ket{n}\bra{n}\Gamma = \mathds{1}.
\end{align}
In the eigenbasis, the displacement \(\ket{x}\) can be expressed as,
\begin{align} \label{x_in_eigenbasis}
    \ket{x} = \sum_{n=1}^N y_n \ket{n}, \quad y_n  := \dfrac{\braket{n|\Gamma|x}}{\braket{n|\Gamma| n}} = \dfrac{1}{\gamma_n} \braket{n|\Gamma|x} = x_{n} - x_{n+1}.
\end{align}
\( \{ y_n \} \) correspond to the eigenmode amplitudes or simply, eigenmodes, as projections onto the eigenvectors of \(\mathrm{W}\). Equation of motion in Eq.~\eqref{explicit_EOM} decouples in the eigenbasis as eigenmode \(y_n\) evolves via, 
\begin{align} \label{EOM_y}
    \dot{y}_n  + \lambda_n y_n = \dfrac{1}{\gamma_n} \braket{n| \Gamma \Gamma^{-1} |f} = \dfrac{1}{\gamma_n} \braket{n|f};  
\end{align}
we introduce the following notation for the transformed force,
\begin{align} \label{modified_force}
    F_n = \braket{n|f} =   \sum_{i=1}^n f_i .
\end{align} 
Thus, the force \(F_n\) for eigenmode $n$  is a sum of forces $f_i$ from $1$ to $n$. Finally, we rewrite Eq.~\eqref{EOM_y},
\begin{align} \label{KV_EOM_eigenbasis}
    {\gamma_n} \dot{y}_n + {\kappa_n} y_n = F_n,
\end{align}
with the solution, 
\begin{align} \label{solution_eigenbasis}
    %y_i(t) =  e^{- \frac{\kappa_i}{\gamma_i}(t-t_0)} y_i(t_0) + \dfrac{1}{\gamma_i} \sum_{j=1}^i \int_{t_0}^t d t^{\prime}  e^{-\frac{\kappa_i}{\gamma_i}(t - t^{\prime})} f_j (t^{\prime})
     y_n(t) =  e^{- \frac{\kappa_n}{\gamma_n}(t-t_0)} y_n(t_0) +  \int_{t_0}^t d t^{\prime}  e^{-\frac{\kappa_n}{\gamma_n}(t - t^{\prime})} \dfrac{F_n (t^{\prime})}{\gamma_n}.
\end{align}
Eq.~\eqref{solution_eigenbasis} will be used in finding the systems susceptibility in the coming section.

\begin{figure*} 
    \centering
    \includegraphics[width=\linewidth]{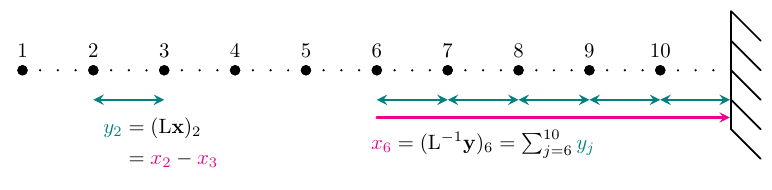}
    \caption{\justifying \textbf{Schematic for eigenmodes}, exemplarily for  \(N = 10\). Left part of sketch demonstrates the eigenmode  \(i = 2\) via application of $\mathrm{L}$, see  Eq.~\eqref{normal_mode_disp}. 
    Right part of sketch illustrates action of  $\mathrm{L}^{-1}$ to obtain $x_6$ from eigenmodes, 
   see Eq.~\eqref{disp_normal_mode}.}
    \label{fig_normalmodes}
\end{figure*}

Using the definition in Eq.~\eqref{x_in_eigenbasis}, one constructs \( \mathbf{y} = (y_1,\dots, y_N)^\top \), and finds from Eq.~\eqref{L_matrix},
%The eigenmodes $\mathbf{y}$  are given by,  
\begin{align} \label{normal_mode_disp}
   \mathbf{y} = \mathrm{L} \mathbf{x}; \quad  y_i = x_i - x_{i+1}.
\end{align}
As anticipated, the eigenmodes correspond to the bonds, i.e., to the relative displacements between two neighboring DoFs (see Fig.~\ref{fig_normalmodes}), whose  dynamics is thus decoupled.  
%The boundary condition at the edge is ensured by setting, $x_{M+1} = 0$.S^{-1} \bm{\zeta} \equiv
The displacements $\mathbf{x}$ are found from the eigenmodes $\mathbf{y}$ by inversion of Eq.~\eqref{normal_mode_disp},
\begin{align} \label{disp_normal_mode}
    \mathbf{x} = \mathrm{L}^{-1} \mathbf{y}; \quad x_i = \sum^N_{j=i} y_j.
\end{align}
Hence, \(\mathrm{L}^{-1}\) acts as a cumulative sum  (discrete integration) operator with elements, 
\begin{align} \label{LinvS_matrix}
    (\mathrm{L}^{-1})_{ij} = 
    \begin{cases}
    1, \quad i \le j, \\ 
    0, \quad i > j;
    \end{cases} \quad (\mathrm{L}^{-1} \mathbf{x})_i = \sum_{j=i}^N x_j.
\end{align}

\section{Response: ``X-ray vision" and infinite range} \label{cor_res}
We aim to find the linear response of \( \mathbf{x}  (t) \) to the application of a time-dependent force vector $\mathbf{f}(t)$,
\begin{align} \label{NE_linear_response}
     \mathbf{x}  (t) =  \int_{-\infty}^t dt^{\prime} \chi (t-t^{\prime}) \mathbf{f}(t^{\prime}),
\end{align}
with  the susceptibility $\chi (t) \in \mathbb{R}^{N \times N} $, formally defined as,
\begin{align} \label{susceptibility}
    \chi_{ij} (t - t^{\prime} ) =  \left. \dfrac{\delta  x_i  (t)}{\delta f_j (t^{\prime}) } \right |_{\mathbf{f} = \mathbf{0}} .
\end{align}
\(\chi_{ij} (t)\) thus quantifies the response at site $i$ due to a force applied at site $j$. $\chi (t) $ can be read off from Eq.~\eqref{MOUP_solution}, letting \(t_0 \to -\infty\), 
\begin{align}
    \chi (t) = \Theta (t) \exp{(- \mathrm{W} t)} \Gamma^{-1}.
\end{align}
\( \Theta(t) \) is the Heaviside step function owing to causality. We may further express $\chi (t)$ in the diagonal representation,  
\begin{align} \label{chi_diagonal}
    \chi (t) = \Theta (t) \mathrm{L}^{-1} e^{- \Lambda {t} } \Lambda_\Gamma^{-1} (\mathrm{L}^{-1})^\top \mbox{, i.e., } \chi_{ij}(t) = \Theta (t)  \mathrm{L}^{-1}_{il} \left ( \dfrac{1}{\gamma_l}e^{-\frac{\kappa_l}{\gamma_l} t} \right ) \mathrm{L}^{-1}_{jl}. 
\end{align}
We used, \( (\mathrm{L}^{-1})^\top_{lj}= \mathrm{L}^{-1}_{jl} \), and adopted Einstein summation convention where summation over repeated indices is implied. Plugging \(\mathrm{L}^{-1}\) from Eq.~\eqref{LinvS_matrix} in Eq.~\eqref{chi_diagonal}, \(\chi_{ij} (t)\) can be expressed as,
\begin{align} \label{chi_ij_clean}
    \chi_{ij} (t) = \Theta (t) \sum_{l = \max{\{i,j\}}}^N \dfrac{1}{\gamma_l}e^{-\frac{\kappa_l}{\gamma_l} t}.
\end{align}
\begin{figure*} 
    \centering
    \includegraphics[width=\linewidth]{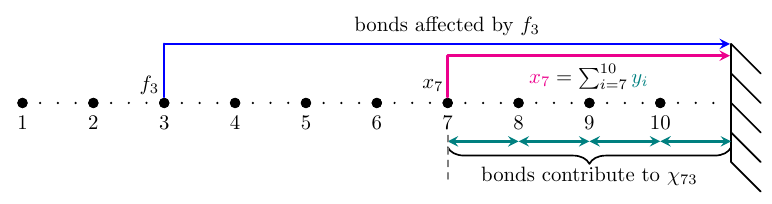}
    \caption{\justifying \textbf{Schematic for the susceptibility} \(\chi_{ij}\) in Eq.~\eqref{chi_ij_clean}, quantifying  the response at  site \(i\) to a force at  site \(j\), exemplarily for \(N = 10\). A force acting on \(j = 3\) yields, via Eq.~\eqref{modified_force}, a finite force $F_l$ for \(l \ge 3\), so that these bonds show a response to the force. On the other hand, the response at  site \(i=7\) is found from the   bonds with \(l \ge 7\), see Eq.~\eqref{disp_normal_mode}. Hence, the susceptibility \(\chi_{73}=\chi_{37}\) has contributions from bonds \(l \ge \max\{7,3\}\), i.e., \(l \ge 7\). }
    \label{fig_response}
\end{figure*}
This expression has an intriguing and transparent origin (see Fig.~\ref{fig_response}):
\begin{enumerate}
    \item A  force $f_j$ applied at site $j$ yields  a  force $F_l$ from Eq.~\eqref{modified_force},
    \begin{align}
        F_l= f_j \Theta(l-j) =\left\{\begin{array}{cc}
            0, & l<j; \\
            f_j, & l\geq j.
        \end{array}\right.
    \end{align}
    In other words, the force $f_j$ only affects eigenmodes with $l\geq j$.
    \item  The displacement (response) at site $i$ is obtained by summing over  eigenmodes $l$ with $l \ge i$,  see Eq.~\eqref{disp_normal_mode}. 
    \item As a result of these two observations, the response $\chi_{ij}(t)$ involves only those eigenmodes that lie in the overlap of these two regions, namely $l \ge \max\{i,j\}$, as indicated in Eq.~\eqref{chi_ij_clean}.
\end{enumerate}

\begin{figure*}
    \centering
    \includegraphics[width=0.6\linewidth]{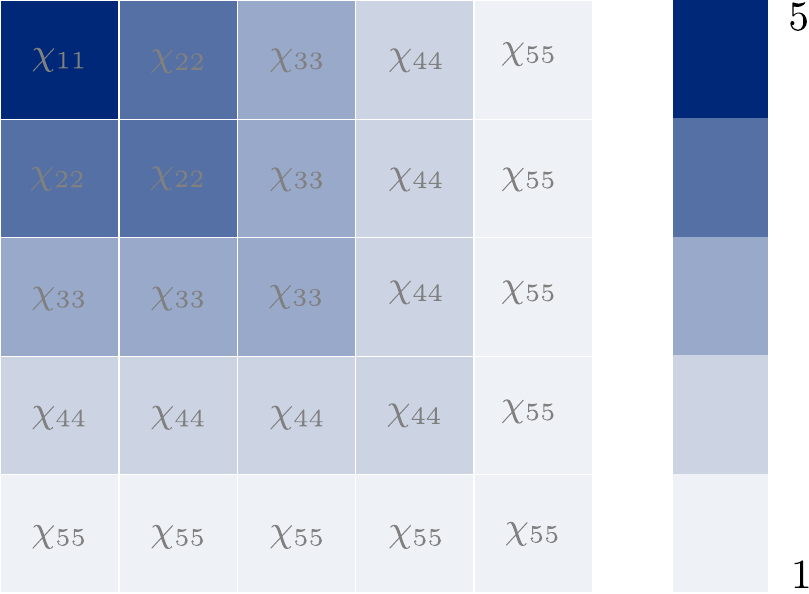}
    \caption{\justifying \textbf{Susceptibility matrix}, exemplary for \(N = 5\). The color-code gives the number of bonds contributing to the respective element \(\chi_{ij}\) according to Eq.~\eqref{chi_ij_clean}. % \(\chi_{ij}\) contains at most \(N = 5\) terms as in \(\chi_{11}\), or at least \(1\) element as in \(\chi_{55}\). 
    Any element  \(\chi_{ij}\)  equals a diagonal entry, i.e., $\chi_{ij} = \chi_{ji} = \chi_{ll}$ with $l= \max{\{i,j\}}$. }
    \label{fig_response_matrix}
\end{figure*}

The susceptibility thus naturally reflects the cumulative nature of both force transmission and displacement reconstruction, leading to the compact structure in Eq.~\eqref{chi_ij_clean}. The susceptibility in Eq.~\eqref{chi_ij_clean} has the following physical properties that are worth noting:
\begin{enumerate}
    \item ``X-ray vision": The response of site $i$ to a force acting on site $j$ is {\it independent} of the properties of sites with $l < \max{\{i,j\}}$. It is especially independent of the properties of the sites that lie in between $i$ and $j$, i.e., the response \emph{sees through} the sites in between $i$ and $j$. As an extreme example: The response of site $N$ to a force acting on site 1 depends only on $\kappa_N$ and $\gamma_N$. As the susceptibility is symmetric, this also applies to the response of site $1$ to a force acting on site $N$.  \label{rontgen}
    \item Infinite range: The response is of infinite range as  it does not depend on the number of sites located between $i$ and $j$.   \label{Unendlichkeit}
    \item Any off diagonal entry equals a diagonal one, i.e., $\chi_{ij}(t) = \chi_{ll}(t)$ with $l= \max{\{i,j\}}$. As a consequence, the susceptibility matrix may at most contain $N$ different entries (see Fig.~\ref{fig_response_matrix}). \label{Treppe}
    \item The response $\chi_{ii}$ does not depend on the order of the bonds with $l\geq i$. As an extreme example, $\chi_{11}$, i.e., the response of site 1 to a force acting on site 1, does not depend on the order of the bonds in the chain. \label{blind}
\end{enumerate}

\section{Rank-deficient interaction: Two sets of distinct modes} \label{sp_case}
\subsection{Constrained and free modes}
An interesting example concerns the case of a rank-deficient interaction matrix, as it leads to distinct types of response. We  may partition the set of bonds, $\mathcal{B} = \{1,\dots,N\}$, into two subsets: constrained bonds, $i \in \mathcal{B}_\kappa$, correspond to nonzero stiffness coefficients, $\kappa_i > 0$, while free bonds, $i \in \mathcal{B}_0$, correspond to $\kappa_i = 0$, with $\mathcal{B}_0 = \mathcal{B} \setminus \mathcal{B}_\kappa$,
\begin{equation}
\kappa_i =
\begin{cases}
> 0, & i \in \mathcal{B}_\kappa, \\
0, & i \in \mathcal{B}_0.
\end{cases}
\end{equation}
The presence of free bonds renders the stiffness matrix $\mathrm{K}$ rank-deficient and partitions the chain into disconnected segments. Each such segment supports a mode responding to rigid translation without restoring force. Consequently, the spectrum of the drift operator $\mathrm{W}$ contains zero eigenvalues (cf.~Eq.~\eqref{eigenspectrum}), and the state space decomposes into two physically distinct subspaces: the null space $\operatorname{null}(\mathrm{W})$, spanned by the modes with $\kappa_i=0$, and a complementary subspace, the image space $\operatorname{im}(\mathrm{W})$, spanned by the modes with finite $\kappa_i$, 
\begin{equation}
\mathbb{R}^N = \operatorname{null}(\mathrm{W}) \oplus \operatorname{im}(\mathrm{W}).
\end{equation}
%Thus, cut bonds introduce unconstrained, diffusive degrees of freedom, while intact bonds determine the spectrum and structure of the relaxing modes.
This distinction is directly reflected in the dynamics. From Eqs.~\eqref{EOM_y}, modes associated with nonzero eigenvalues relax exponentially in time due to restoring forces (\(\kappa_i \ne  0\)), whereas the free modes, in absence of external forces, do not evolve. The dynamics therefore separates naturally into relaxing (constrained) and steady (unconstrained) components.

The linear response of the system in Eq.~\eqref{chi_ij_clean} can in this case be written as,
\begin{align} \label{chi_ij_rank_deficient}
    %\chi_{ij} (t) = \Theta (t) \left( \sum_{l \in \mathcal{B}_0} \dfrac{1}{\gamma_l} + \sum_{l \in \mathcal{B}_\kappa} \dfrac{1}{\gamma_l}e^{-\frac{\kappa_l}{\gamma_l} t} \right )
    \chi_{ij}(t) = \Theta(t) \left(
    \sum_{\substack{l \in \mathcal{B}_0 \\ l \ge \max\{i,j\}}} \frac{1}{\gamma_l}
    + \sum_{\substack{l \in \mathcal{B}_\kappa \\ l \ge \max\{i,j\}}} \dfrac{1}{\gamma_l}e^{-\frac{\kappa_l}{\gamma_l} t}
    \right).
\end{align}
The susceptibility thus naturally decomposes into two contributions associated with distinct dynamical subspaces. Free modes correspond to $\mathcal{B}_0$ (equivalently, $\operatorname{null}(\mathrm{W})$) as they respond instantaneously to external perturbations, producing a time-independent purely viscous contribution. In contrast, constrained modes are associated with $\mathcal{B}_\kappa$ (equivalently, $\operatorname{im}(\mathrm{W})$), and they exhibit exponentially decaying responses, reflecting their  elastic component.

In the two next subsections, we will  elucidate driving protocols that exclusively probe one of the two subspaces.  

\begin{figure*} 
    \centering
    \includegraphics[width=\linewidth]{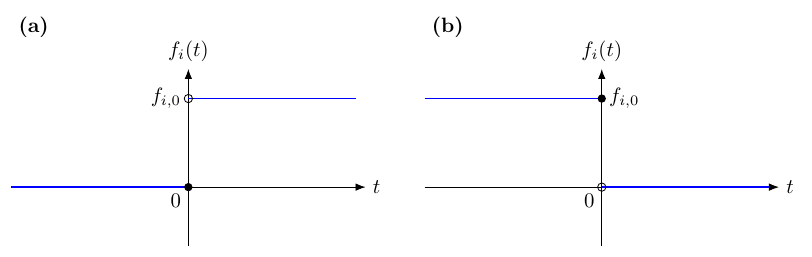}
    \caption{\justifying \textbf{Protocols designed to specifically probe the free and constrained modes}. (a) ``Steady driving" as depicted in Eq.~\eqref{steady_protocol} where force acts only at \(t > 0\), revealing the free modes. (b) ``Relaxation" as depicted in Eq.~\eqref{single_step_protocol}, where force only acts at \(t \le 0\),  unveiling the constrained modes for $t>0$.}
    \label{fig_protocols}
\end{figure*}

\subsection{Steady driving: Probing the free modes}
We start with the case of a steady driving, i.e., 
\begin{align} \label{steady_protocol}
   f_{i}(t) &= \begin{cases}
                0 &  t \le 0, \\
                f_{i,0} & t > 0. \\
            \end{cases} 
\end{align}
In this case, the solution for \(x_i (t)\) from Eq.~\eqref{MOUP_solution} is given by, 
\begin{align} \label{xi_driving}
    {x}_{i} (t)=  \Theta (t)\left(
    \sum_{\substack{l \in \mathcal{B}_0 \\ l \ge i}} \dfrac{F_{l,0} t}{\gamma_l}
    + \sum_{\substack{ l \in \mathcal{B}_\kappa \\ l \ge i }} \dfrac{F_{l,0}}{\kappa_l} (1 - e^{-\frac{\kappa_l}{\gamma_l} t})
    \right); 
\end{align}
with \(F_{l,0} = \sum_{j=1}^l f_{j,0}\), as in Eq.~\eqref{modified_force}. The steady state is reached when all modes with finite $\kappa$ have relaxed, and the steady state velocity \(\dot{x}_i\) is completely determined by the free modes,
\begin{align}
    \lim_{t \to \infty} \dot{x}_i (t) = \sum_{\substack{l \in \mathcal{B}_0 \\ l \ge i}} \dfrac{F_{l,0} }{\gamma_l}.
\end{align}

\subsection{Relaxation: Probing the constrained modes}
We consider in this section the opposite protocol scenario to the above, namely,
\begin{align} \label{single_step_protocol}
   f_{i}(t) &= \begin{cases}
                                  f_{i,0} &  t \le 0, \\
                                 0 & t > 0. \\
                            \end{cases} 
\end{align}
The force applied during negative times gives rise to an initial condition $\mathbf{x}(t=0)$. The solution for $t>0$ is then found from Eq.~\eqref{MOUP_solution}, by setting \( \mathbf{f} = \mathbf{0} \),
\begin{align} 
    \mathbf{x} (t) & = e^{- \mathrm{W} t}\mathbf{x} (0) ,\\
    \dot{\mathbf{x}} (t) 
    &= -\mathrm{W}e^{- \mathrm{W} t}\mathbf{x} (0).
\end{align}
For the eigenmodes, 
\begin{align}
    \mathbf{y} (t) &=  e^{- \Lambda t}   \mathbf{y}  (0) ,\\
     \dot{\mathbf{y}} (t) &=  -\Lambda e^{- \Lambda t}   \mathbf{y}  (0).\label{eq:recoilnormal}
\end{align}
with the diagonal matrix $\Lambda$ given in Eq.~\eqref{eigenspectrum}. Eq.~\eqref{eq:recoilnormal} shows that only the non-zero eigenvalues contribute to motion for positive times,
\begin{align} \label{odd_recoil_single-step}
   \dot y_{i} (t>0) = \begin{cases}
        0, & \forall i \in \mathcal{B}_0 ; \\
         -\dfrac{F_{i,0}}{\gamma_i} e^{-\frac{\kappa_{i}}{\gamma_{i}}t}, & \forall i \in \mathcal{B}_\kappa. 
    \end{cases} 
\end{align}
As before, \( F_{i,0} = \sum_{j=1}^i f_{j,0} \). %The initial value \(y_i (0)\) for a slow mode is found by putting $t = t_\text{on}$, in Eq.~\eqref{xi_driving}.
The negative sign signifies that the relaxation motion (recoil) is in the direction opposite to  driving $f_{i,0}$ \cite{Ginot_2022, Caspers_2023, Cao2023, Vaidya_2025}. For the DoF $x_i$, we finally have for  $t>0$,%The recoil of any DoF can now be calculated using Eq.~\eqref{disp_normal_mode},
\begin{align} \label{recoil_object_single-step}
\begin{split}
         \dot x_i (t) &= \sum_{\substack{ l \ge i \\ l \in \mathcal{B}_\kappa}} \dot y_{l} (t) 
         = -  \sum_{\substack{ l \ge i \\ l \in \mathcal{B}_\kappa}} \dfrac{F_{l,0}}{\gamma_l} e^{-\frac{\kappa_{l}}{\gamma_{l}} t}.
    \end{split}
\end{align}

This shows that the relaxation dynamics is solely governed by the constrained modes. 

In summary, when the force that has been acting for a long time is switched off, the motion of all sites instantaneously changes from being completely governed by the free modes to being completely governed by the constrained modes. Such protocol thus enables detection of the two sets of modes individually.

\section{Example: Viscoelastic fluid in confinement} \label{example}

\begin{figure*} 
    \centering
    \includegraphics[width=0.6\linewidth]{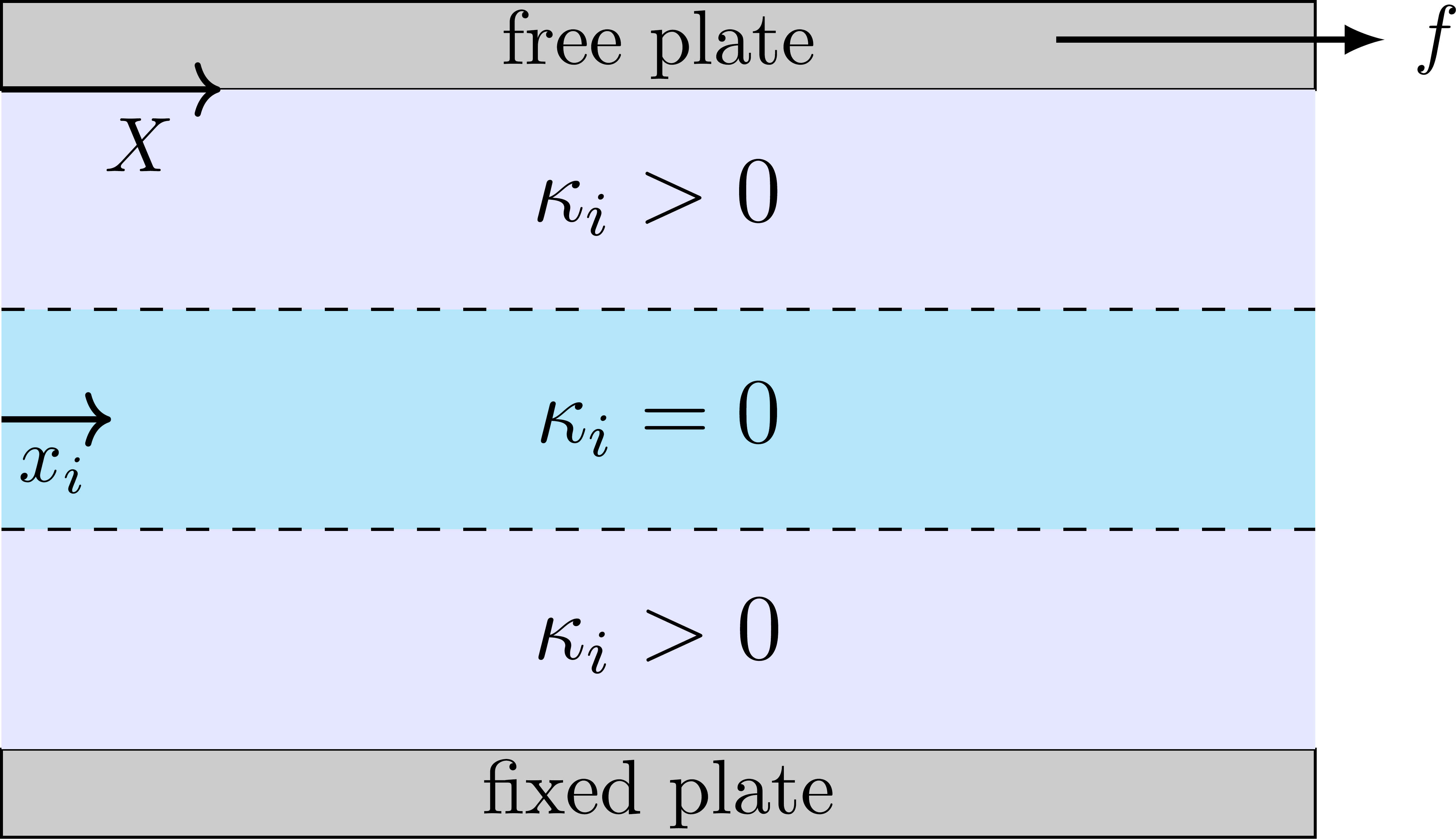}
    \caption{\justifying \textbf{Viscoelastic fluid confined in-between two parallel plates}. The lower plate is fixed, while the upper plate is subject to a force $f$. The fluid is divided into three regions: viscoelastic regions (purple) located near both plates, characterized by \(\kappa_i >  0\), and a purely viscous region (sky blue) sandwiched between them, where \(\kappa_i =  0\). When the upper plate is driven, the layers of the fluid are sheared. The lateral displacement of layer \(i\) is denoted by \(x_i\), while the displacement of the upper plate is denoted by \(X\).}
    \label{fig_simple_shear}
\end{figure*}

We aim to illustrate the above findings in an example system of a confined  fluid. 
Consider  a simple fluid  confined between two parallel plates. The lower plate is kept fixed, while the upper plate is free to move laterally, and subject to a time dependent force $f(t)$.
The coordinates $x_i$ correspond  to lateral displacements of the corresponding fluid layer at height indexed by $i$, with $x_1\equiv X$, the position of the upper plate. The vicinities of the plates  possess finite restoring forces, e.g., corresponding to polymer brushes attached to the surfaces of the plates. This is mimicked by finite values of $\kappa_i$ close to the two plates. The simple fluid in between the plates is modelled by $\kappa_i=0$ (see Fig.~\ref{fig_simple_shear}). The following statements hold true for any positive values of $\gamma_i$ and any non-negative $\kappa_i$.    

%In this setting, the following phenomena occur:
\begin{itemize}
\item Under a time independent force $ f $, the steady state velocity of the upper plate only depends on the bulk region, i.e., only on the free modes with $\kappa_i=0$,
\begin{align}
    \lim_{t \to \infty} \dot{X} (t) = f\sum_{l \in \mathcal{B}_0} \dfrac{1}{\gamma_l}.\label{eq:response1}
\end{align}
\item After switching off the force, the recoil motion of the upper plate is solely determined by the constrained modes  with $\kappa_i\not=0$, 
\begin{align}
         \dot X (t) = - f  \sum_{ l \in \mathcal{B}_\kappa} \dfrac{1}{\gamma_l} e^{-\frac{\kappa_{l}}{\gamma_{l}} t}.\label{eq:response2}
    \end{align}
    \item The above statements, i.e., Eqs.~\eqref{eq:response1} and  \eqref{eq:response2}, do not depend on the order of the segments. For example, exchanging the positions of the three regions in Fig.~\ref{fig_simple_shear} has no effect on the response of $X$. 
    \item The response of  any layer to the force $f$ applied to the upper plate is independent of the material above the respective layer. For example, the response of the bulk fluid in the center is independent of the polymer layer attached to the upper plate.
\end{itemize}

This example provides a direct physical realization of the decomposition into free and constrained modes: steady driving probes the free (unconstrained) subspace, whereas transient relaxation reveals the constrained subspace. 

\section{Maxwell chain as a singular limiting case} \label{maxwell_limit}
\begin{figure*} 
    \centering
    \includegraphics[width=0.75\linewidth]{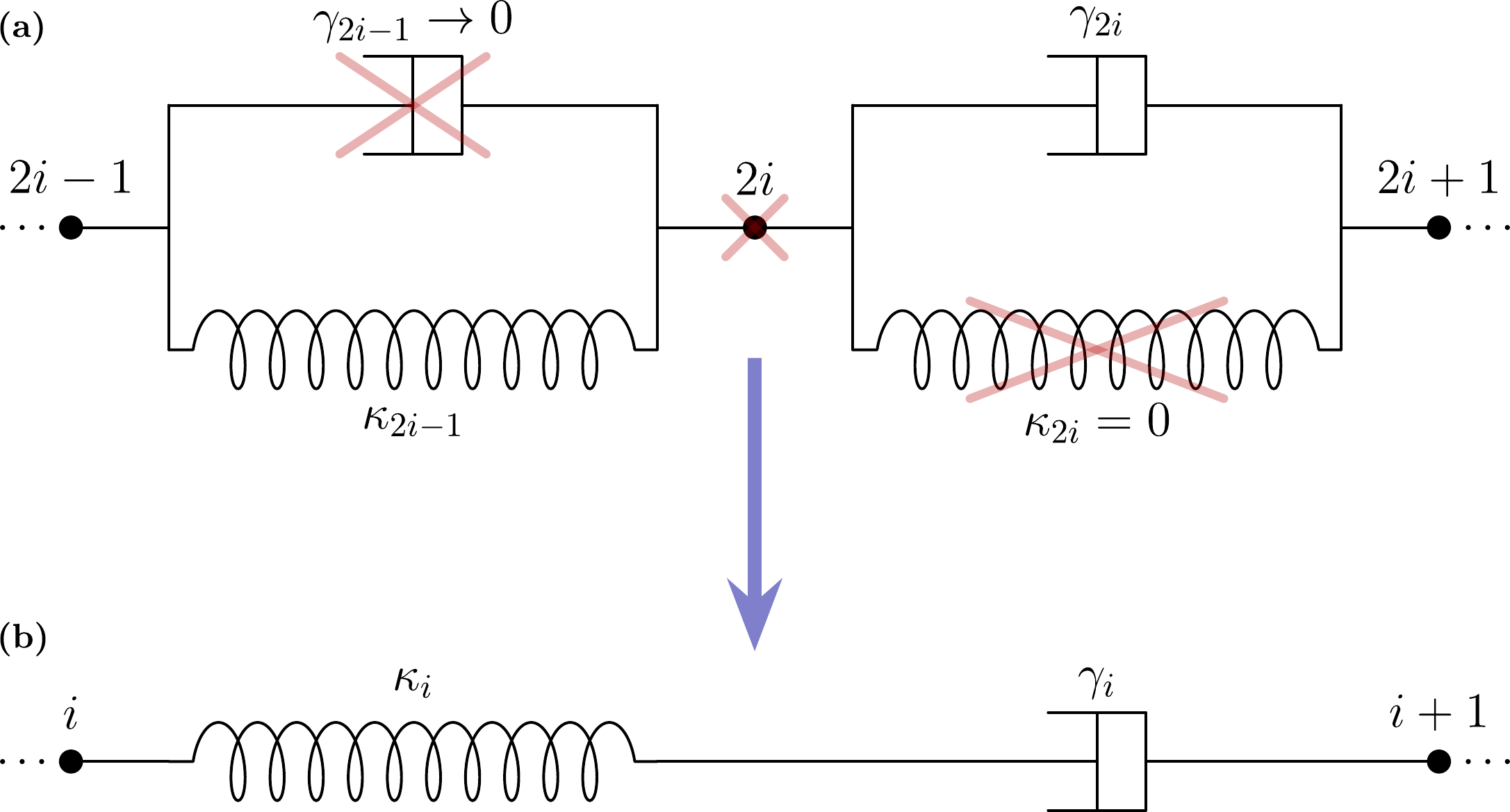}
    \caption{\justifying \textbf{Construction of a Maxwell chain as a singular limit of the Kelvin--Voigt chain with even number of bonds.}
    %(a) Neighboring bonds $(2i-1,2i)$ in a Kelvin--Voigt chain are grouped into pairs. 
    (a) Setting $\kappa_{2i}=0$ removes every second spring, while taking the singular limit $\gamma_{2i-1}\to0$ suppresses the complementary dashpots. The intermediate node is then eliminated.
    (b) Each pair of bonds reduces to a Maxwell element, comprising a spring of stiffness $\kappa_i$ in series with a dashpot of friction coefficient $\gamma_i$. This yields a chain of Maxwell elements.}
    \label{fig_Maxwell}
\end{figure*}

The Kelvin--Voigt chain considered above also contains Maxwell chains as limiting topologies, see Fig.~\ref{fig_Maxwell}. 
To illustrate this, we consider a Kelvin--Voigt chain with an even number of bonds.
From it, the springs on every second bond are cut (removed), by setting $\kappa_{2i} =0$. 
A Maxwell chain is then obtained by additionally removing the dashpot in every other bond, i.e., by taking the limit $\gamma_{2i-1} \to 0^{+}$. 
%Each pair of bonds then reduces to an elastic spring in series with a dashpot. 
This is, however, a singular limit of the present overdamped formulation, since the friction matrix \(\Gamma\) ceases to be positive definite with  $\gamma_{2i-1} =0$. Performing this limit carefully in Eq.~\eqref{chi_ij_clean} turns $\frac{1}{\gamma} e^{-\frac{\kappa}{\gamma} t}$ into a delta distribution~\footnote{ \( \lim_{\gamma\to0^+} \frac{1}{\gamma} e^{-\frac{\kappa}{\gamma} t} \Theta(t) = \frac{1}{\kappa}\delta(t)\), where  $\delta(t)$ is the one-sided delta distribution: \(\int_{-\infty}^{t}dt'\,\delta(t-t')x(t')=x(t)\)}. 

To compare the response between Kelvin--Voigt and Maxwell chains, we eliminate the intermediate (massless) nodes   and relabel the remaining ones such that nodes \(i\) and \(i+1\) are connected by spring \(\kappa_i\) and a dashpot \(\gamma_i\) in series, see Fig.~\ref{fig_Maxwell}.
This yields the  susceptibility \(\chi^{\rm M}_{ij}\) in a Maxwell-chain, 
\begin{equation}
    \chi^{\rm M}_{ij}(t) = \sum_{l=\max\{i,j\}}^{N} \left[ \frac{1}{\kappa_l}\delta(t) + \frac{1}{\gamma_l}\Theta(t) \right]. \label{Maxwell_susceptibility}
\end{equation}
Notably, \(\chi^{\rm M}_{ij}\) in Eq.~\eqref{Maxwell_susceptibility} has the same staircase structure of the response, including its ``infinite-range'' and ``X-ray vision'' properties discussed in Sec.~\ref{cor_res} as the Kelvin--Voigt susceptibility in Eq.~\eqref{chi_ij_clean}. What differs is the single-element response kernel: a Maxwell element, consisting of a spring and dashpot in series, gives an instantaneous elastic response $\kappa_l^{-1}\delta(t)$, and a viscous one $\gamma_l^{-1}\Theta(t)$ while a Kelvin--Voigt element gives an exponential relaxation kernel \(  \frac{1}{\gamma_l} e^{-\frac{\kappa_l}{\gamma_l}t} \Theta(t)\). 

\section{Relation to Fluctuations} \label{FDT}
In this section, we relate the response of the system to its equilibrium fluctuations at inverse temperature $\beta=(k_{\rm B}T)^{-1}$.
The Fluctuation-Dissipation Theorem (FDT) \cite{Kubo1966FDT}
%Onsager's regression hypothesis \cite{Onsager1931One, Onsager1931Two} 
connects the linear response  to spontaneous fluctuations \cite{hansen2013theory},
\begin{align} \label{MOUP_response_correlation}
    \chi (t) =  -\beta \Theta (t) \dfrac{d}{dt} \mathcal{C}(t).
\end{align}
In Eq.~\eqref{MOUP_response_correlation}, %s a direct manifestation of the Fluctuation-Dissipation Theorem (FDT) \cite{Kubo1966FDT}, where 
\(\chi(t)\) is the susceptibility matrix defined in Eqs.~\eqref{NE_linear_response},~\eqref{susceptibility}, and \( \mathcal{C}(t) \) is the equilibrium correlation matrix in absence of force, describing spontaneous fluctuations in the system. 
\(\mathcal{C}(t)\) is defined as follows \cite{godreche2018characterising},
\begin{align} \label{MOUP_Correlation}
    \mathcal{C}(t-t^{\prime}) = \langle \mathbf{x} (t) \mathbf{x} ^\top (t^{\prime})\rangle = \exp{(-\mathrm{W} \abs{t-t^{\prime}})} \mathcal{C} (0).
\end{align}
\( \mathcal{C} (0) = \langle \mathbf{x} \mathbf{x}^\top  \rangle = \beta^{-1}\mathrm{K}^{-1}\) is the covariance matrix \cite{risken1996fokker}. 
In the diagonal representation, $\mathcal{C}(t)$ is expressed as,   
\begin{align} \label{definition_correlation_matrix}
    \beta\mathcal{C}(t) = \mathrm{L}^{-1} e^{- \Lambda \abs{t} } \Lambda_\mathrm{K}^{-1} (\mathrm{L}^{-1})^\top \mbox{, i.e., } \beta \mathcal{C}_{ij}(t) = \mathrm{L}^{-1}_{il} \left ( \dfrac{1}{\kappa_l} e^{-\frac{\kappa_l}{\gamma_l}\abs{t} } \right) \mathrm{L}^{-1}_{jl}; 
\end{align}
We used, \( (\mathrm{L}^{-1})^\top_{lj}= \mathrm{L}^{-1}_{jl} \), and adopted Einstein summation convention as before. Plugging \(\mathrm{L}^{-1}\) from Eq.~\eqref{LinvS_matrix} in Eq.~\eqref{definition_correlation_matrix}, \(\mathcal{C}_{ij}(t)\) can be expressed as,
\begin{align} \label{c_ij}
    \beta \mathcal{C}_{ij}(t) = \sum_{l = \max{\{i,j\}}}^N \dfrac{1}{\kappa_l} e^{-\frac{\kappa_l}{\gamma_l} \abs{t}}.
\end{align}
The susceptibility \(\chi_{ij}\) may now be evaluated using Eq.~\eqref{MOUP_response_correlation}, which recovers Eq.~\eqref{chi_ij_clean}. 

Importantly, the existence of $\mathrm{K}^{-1}$ is required for the correlation matrix to be finite. If some $\kappa_l=0$, i.e., if free modes exist (compare Sec.~\ref{sp_case}), the corresponding contribution to $C_{ij}(t)$ diverges. The susceptibility, however, remains finite because it is related to the time derivative of the correlation,  %For %$t>0$,
\begin{equation}
   -\frac{d}{dt} \left \{ 
   \frac{1}{\kappa_l}
    \exp\left(-\frac{\kappa_l}{\gamma_l}t\right) \right \}
        =
   \frac{1}{\gamma_l}
    \exp\left(-\frac{\kappa_l}{\gamma_l}t\right)
     \xrightarrow{\kappa_l \to 0^+}
   \frac{1}{\gamma_l} > 0 .
\end{equation}
Therefore, for rank-deficient $\mathrm{K}$, the FDT expression should be interpreted as a regularized limit $\kappa_l\to0^+$.

\section{Conclusion} \label{con}

We developed an exact analytical framework for a class of  heterogeneous overdamped chains with harmonic interactions and momentum-conserving dissipation. Despite the non-symmetric nature of the drift operator $\mathrm{W} = \Gamma^{-1} \mathrm{K}$, we demonstrate that its structure admits an exact diagonalization through a forward-difference transformation, mapping the dynamics from node to bond variables. This holds in the presence  of heterogeneity in terms of interactions or damping.

A central result is the emergence of a staircase structure in the linear response function as illustrated in Figs.~\ref{fig_response},~\ref{fig_response_matrix}. The susceptibility $\chi_{ij}$ depends only on modes $l$ with $l \geq \max\{i,j\}$, leading to the striking consequence that the response between two sites is independent of the properties of the system in between them. This ``X-ray vision'' property implies an effectively infinite-range response and a highly constrained structure of the susceptibility matrix, with at most $N$ distinct entries.

We further showed that when the interaction matrix is rank-deficient, the dynamics naturally decomposes into two physically distinct subspaces. Free modes, associated with zero eigenvalues, produce an instantaneous, purely viscous response and the steady-state behavior under sustained driving is exclusively governed by these modes. In contrast, the transient relaxation dynamics after removal of forcing, is exclusively  governed by the constrained modes, associated with finite eigenvalues. This separation provides a way of individual probing of the two types of modes via choice of protocol. %transparent dynamical interpretation of viscoelastic behavior within a purely overdamped setting, without invoking inertia.

The mentioned properties were illustrated in the example of a fluid under shear. Beyond the simple-shear geometry considered here, the present framework can be extended to a broader class of flow configurations, including other shear geometries such as Taylor–Couette flow \cite{Taylor1923, landau1987fluid}. More generally, the model also provides a natural description of the response of rotating objects embedded in viscoelastic fluids, where the perturbation is generated not by an external force but by an applied torque acting again at a \emph{free end}. This opens a route toward analytically characterizing torsional microrheology and torque-driven viscoelastic response within the same unified framework \cite{Niloy2026}.

Furthermore, Maxwell chains were obtained as singular limits of the Kelvin--Voigt chain, retaining the same staircase response but replacing the exponential Kelvin--Voigt kernel by an instantaneous elastic response plus a viscous response.

Our results highlight how momentum-conserving dissipation fundamentally alters the structure of overdamped dynamics. Rather than destroying analytical tractability, the combination of relative friction and nearest-neighbor interactions imposes a structure that enables an exact solution, even in heterogeneous systems. This stands in contrast to conventional overdamped models with local friction, where heterogeneity typically precludes closed-form analysis.

The framework developed here opens several directions for future research. A natural extension is to incorporate stochastic forces and investigate fluctuation properties beyond the equilibrium setting discussed here, in particular when the stochastic forces do not obey the fluctuation--dissipation theorem.  Another promising direction is the study of time-dependent or disordered network topologies, where bonds may dynamically appear or disappear, potentially leading to non-trivial aging or memory effects. Non-reciprocal bonds are also of interest \cite{Brandenbourger2019, Scheibner2020OddElasticity}. Extending the analysis to higher-dimensional networks or more complex geometries could reveal how the ``X-ray'' response generalizes beyond one dimension. Such extensions would naturally connect to mechanical models of protein allostery, where one studies how a force or perturbation applied at one site affects the conformation at a distant site~\cite{Tsvi_Tlusty2018Protein, RevModPhys.91.031001}, characterized by the Green function, analogous to the susceptibility $\chi_{ij}$ considered here. Finally, introducing nonlinear interactions or coupling to external flows may provide a route to understanding emergent viscoelastic behavior in driven soft-matter systems and complex fluids.

\section*{Acknowledgments} 
This project was funded by the Deutsche Forschungsgemeinschaft (DFG), Grant No. SFB 1432–Project ID 425217212, Project C05. 
We thank Clemens Bechinger and Niloyendu Roy for a critical reading of the manuscript and for useful comments.

\appendix

\section{Explicit Forms} \label{App_Explicit}
The equations of motion for $N$ interacting DoFs, where \(x_i\) corresponds to the (mean) deviation of $i^{\text{th}}$ DoF from its equilibrium position, are given by the following set of coupled linear differential equations,
\begin{align} \label{equation_of_motion}
    \begin{split}
        m_1  \ddot{x}_1 =& -\gamma_1 (\dot{x}_{1} - \dot{x}_{2}) -\kappa_1 (x_{1} - x_{2})  + f_1,  \\
        m_i  \ddot{x}_i = &-\gamma_i (\dot{x}_{i} - \dot{x}_{i+1}) -\gamma_{i-1} (\dot{x}_{i} - \dot{x}_{i-1}) \\
        &-\kappa_i (x_{i} - x_{i+1}) -\kappa_{i-1} (x_{i} - x_{i-1}) + f_i, \mbox{~} i = 2,\dots,N-1; \\
        m_N  \ddot{x}_N = &-\gamma_N \dot{x}_N -\gamma_{N-1} (\dot{x}_N - \dot{x}_{N-1}) -\kappa_N x_N -\kappa_{N-1} (x_N - x_{N-1}) + f_N. 
    \end{split}
\end{align}   
This set of equations are expressed compactly in Eq.~\eqref{explicit_EOM} of the main text, after taking the overdamped limit, i.e., $m \to 0$. 
The friction matrix $\Gamma$ is explicitly given below, 
\begin{align} \label{Gamma_matrix_explicit}
\Gamma = 
\begin{pmatrix}
\gamma_1           & -\gamma_1         & 0                 & \cdots          & 0           & 0 \\
-\gamma_1          & \gamma_1 + \gamma_2 & -\gamma_2       & \ddots          &             & \vdots \\
0                  & -\gamma_2         & \gamma_2 + \gamma_3 & \ddots       & \ddots      & 0 \\
\vdots             & \ddots            & \ddots            & \ddots          & -\gamma_{N-2} & 0 \\
0                  &                   & \ddots            & -\gamma_{N-2} & \gamma_{N-2} + \gamma_{N-1} & -\gamma_{N-1} \\
0                  & \cdots            & 0                 & 0              & -\gamma_{N-1} & \gamma_{N-1} + \gamma_N 
\end{pmatrix} 
\end{align}   
The stiffness matrix \(\mathrm{K} \) is explicitly given below,
\begin{align} \label{K_matrix_explicit}
\mathrm{K} =
\begin{pmatrix}
\kappa_1           & -\kappa_1         & 0                 & \cdots          & 0           & 0 \\
-\kappa_1          & \kappa_1 + \kappa_2 & -\kappa_2       & \ddots          &             & \vdots \\
0                  & -\kappa_2         & \kappa_2 + \kappa_3 & \ddots       & \ddots      & 0 \\
\vdots             & \ddots            & \ddots            & \ddots          & -\kappa_{N-2} & 0 \\
0                  &                   & \ddots            & -\kappa_{N-2} & \kappa_{N-2} + \kappa_{N-1} & -\kappa_{N-1} \\
0                  & \cdots            & 0                 & 0              & -\kappa_{N-1} & \kappa_{N-1} + \kappa_{N}
\end{pmatrix}
\end{align} 
The forward difference matrix $\mathrm{L}$, and its inverse \( \mathrm{L}^{-1} \), representing cumulative sum, are given below, 
\begin{align} \label{fwd_diff_matrix}
\mathrm{L} = 
\begin{pmatrix}
1 & -1 & 0 & 0 & \cdots & 0 \\
0 & 1 & -1 & 0 & \cdots & 0 \\
0 & 0 & 1 & -1 & \cdots & 0 \\
\vdots & \vdots & \ddots & \ddots & \ddots & \vdots \\
0 & 0 & \cdots & 0 & 1 & -1 \\
0 & 0 & \cdots & 0 & 0 & 1
\end{pmatrix}; \quad 
\mathrm{L}^{-1} = 
\begin{pmatrix}
1 & 1 & 1 & \cdots & 1 \\
0 & 1 & 1 & \cdots & 1 \\
0 & 0 & 1 & \cdots & 1 \\
\vdots & \vdots & \vdots & \ddots & 1 \\
0 & 0 & 0 & \cdots & 1
\end{pmatrix}
\end{align}

\bibliographystyle{apsrev4-2}
\bibliography{refs}

@article{Huygens336,
    author = {Bennett, Matthew and Schatz, Michael F. and Rockwood, Heidi and Wiesenfeld, Kurt},
    title = {Huygens's clocks},
    journal = {Proceedings of the Royal Society A: Mathematical, Physical and Engineering Sciences},
    volume = {458},
    number = {2019},
    pages = {563-579},
    year = {2002},
    month = {03},
    issn = {1364-5021},
    doi = {10.1098/rspa.2001.0888},
    url = {https://doi.org/10.1098/rspa.2001.0888}
}

@article{einstein1907,
	author = {Einstein, A.},
	doi = {10.1002/andp.19063270110},
	journal = {Annalen der Physik},
	number = {1},
	pages = {180–190},
	title = {Die Plancksche Theorie der Strahlung und die Theorie der spezifischen Wärme},
	url = {https://onlinelibrary.wiley.com/doi/abs/10.1002/andp.19063270110},
	volume = {327},
	year = {1907}
}

@article{BornKarman1912,
  author  = {Born, Max and von K\'arm\'an, Theodore},
  title   = {{\"Uber Schwingungen in Raumgittern}},
  journal = {Physikalische Zeitschrift},
  volume  = {13},
  pages   = {297--309},
  year    = {1912}
}

@article{debye1912,
author = {Debye, P.},
title = {Zur Theorie der spezifischen Wärmen},
journal = {Annalen der Physik},
volume = {344},
number = {14},
pages = {789-839},
doi = {https://doi.org/10.1002/andp.19123441404},
url = {https://onlinelibrary.wiley.com/doi/abs/10.1002/andp.19123441404},
year = {1912}
}

@article{rouse1953theory,
    author = {Rouse, Prince E., Jr.},
    title = {A Theory of the Linear Viscoelastic Properties of Dilute Solutions of Coiling Polymers},
    journal = {The Journal of Chemical Physics},
    volume = {21},
    number = {7},
    pages = {1272-1280},
    year = {1953},
    month = {07},
    issn = {0021-9606},
    doi = {10.1063/1.1699180},
    url = {https://doi.org/10.1063/1.1699180}
}

@article{Zimm1956,
    author = {Zimm, Bruno H.},
    title = {Dynamics of Polymer Molecules in Dilute Solution: Viscoelasticity, Flow Birefringence and Dielectric Loss},
    journal = {The Journal of Chemical Physics},
    volume = {24},
    number = {2},
    pages = {269-278},
    year = {1956},
    month = {02},
    issn = {0021-9606},
    doi = {10.1063/1.1742462},
    url = {https://doi.org/10.1063/1.1742462}
}

@article{Hung2018,
	author = {Hung, Jui-Hsiang and Mangalara, Jayachandra Hari and Simmons, David S.},
	doi = {10.1021/acs.macromol.8b00135},
	journal = {Macromolecules},
	number = {8},
	pages = {2887–2898},
	title = {Heterogeneous Rouse Model Predicts Polymer Chain Translational Normal Mode Decoupling},
	url = { https://doi.org/10.1021/acs.macromol.8b00135},
	volume = {51},
	year = {2018}
}

@article{Rolls2017,
	author = {Rolls, Edward and Togashi, Yuichi and Erban, Radek},
	doi = {10.1137/16M108700X},
	journal = {Multiscale Modeling \& Simulation},
	number = {4},
	pages = {1672–1693},
	title = {Varying the Resolution of the Rouse Model on Temporal and Spatial Scales: Application to Multiscale Modeling of DNA Dynamics},
	url = { https://doi.org/10.1137/16M108700X},
	volume = {15},
	year = {2017}
}

@article{chen2023,
    author = {Tian, Xiaofei and Chen, Ye and Xu, Xiaolei and Xu, Wen-Sheng and Chen, Jizhong},
    title = {Theory of mobility of inhomogeneous-polymer-grafted particles},
    journal = {The Journal of Chemical Physics},
    volume = {158},
    number = {20},
    pages = {204904},
    year = {2023},
    month = {05},
    issn = {0021-9606},
    doi = {10.1063/5.0153473},
    url = {https://doi.org/10.1063/5.0153473}
}

@article{khatri2007,
	author = {Khatri, Bhavin S. and McLeish, Tom C. B.},
	doi = {10.1021/ma071175x},
	journal = {Macromolecules},
	number = {18},
	pages = {6770–6777},
	title = {Rouse Model with Internal Friction: A Coarse Grained Framework for Single Biopolymer Dynamics},
	url = {https://doi.org/10.1021/ma071175x},
	volume = {40},
	year = {2007}
}

@article{makarov2013,
    author = {Cheng, Ryan R. and Hawk, Alexander T. and Makarov, Dmitrii E.},
    title = {Exploring the role of internal friction in the dynamics of unfolded proteins using simple polymer models},
    journal = {The Journal of Chemical Physics},
    volume = {138},
    number = {7},
    pages = {074112},
    year = {2013},
    month = {02},
    issn = {0021-9606},
    doi = {10.1063/1.4792206},
    url = {https://doi.org/10.1063/1.4792206}
}

@article{rajarshi2021,
    author = {Kailasham, R. and Chakrabarti, Rajarshi and Prakash, J. Ravi},
    title = {Rouse model with fluctuating internal friction},
    journal = {Journal of Rheology},
    volume = {65},
    number = {5},
    pages = {903-923},
    year = {2021},
    month = {09},
    issn = {0148-6055},
    doi = {10.1122/8.0000255},
    url = {https://doi.org/10.1122/8.0000255}
}

@article{Hoogerbrugge_1992,
	author = {P. J. Hoogerbrugge and J. M. V. A. Koelman},
	doi = {10.1209/0295-5075/19/3/001},
	journal = {Europhysics Letters},
	month = {jun},
	number = {3},
	pages = {155},
	publisher = {},
	title = {Simulating Microscopic Hydrodynamic Phenomena with Dissipative Particle Dynamics},
	url = {https://doi.org/10.1209/0295-5075/19/3/001},
	volume = {19},
	year = {1992}
}

@article{Espanol_1995,
	author = {P. Español and P. Warren},
	doi = {10.1209/0295-5075/30/4/001},
	journal = {Europhysics Letters},
	month = {may},
	number = {4},
	pages = {191},
	publisher = {},
	title = {Statistical Mechanics of Dissipative Particle Dynamics},
	url = {https://doi.org/10.1209/0295-5075/30/4/001},
	volume = {30},
	year = {1995}
}

@article{Huang2007,
    author = {Jiang, Wenhua and Huang, Jianhua and Wang, Yongmei and Laradji, Mohamed},
    title = {Hydrodynamic interaction in polymer solutions simulated with dissipative particle dynamics},
    journal = {The Journal of Chemical Physics},
    volume = {126},
    number = {4},
    pages = {044901},
    year = {2007},
    month = {01},
    issn = {0021-9606},
    doi = {10.1063/1.2428307},
    url = {https://doi.org/10.1063/1.2428307}
}

@article{Das2024,
	author = {Das, Debankur and Vink, Richard and Krüger, Matthias},
	doi = {10.1088/1361-648X/ad2b1d},
	journal = {Journal of Physics: Condensed Matter},
	month = {feb},
	number = {21},
	pages = {215707},
	publisher = {IOP Publishing},
	title = {Friction of a driven chain: role of momentum conservation, Goldstone and radiation modes},
	url = {https://doi.org/10.1088/1361-648X/ad2b1d},
	volume = {36},
	year = {2024}
}

@article{Ginot_2022,
	author = {Ginot, Félix and Caspers, Juliana and Reinalter, Luis Frieder and {Krishna Kumar}, Karthika and Krüger, Matthias and Bechinger, Clemens},
	doi = {10.1088/1367-2630/aca8c7},
	journal = {New Journal of Physics},
	month = {dec},
	number = {12},
	pages = {123013},
	publisher = {IOP Publishing},
	title = {Recoil experiments determine the eigenmodes of viscoelastic fluids},
	url = {https://doi.org/10.1088/1367-2630/aca8c7},
	volume = {24},
	year = {2022}
}

@article{BonetAvalos_1997,
	author = {J. Bonet Avalos and A. D. Mackie},
	doi = {10.1209/epl/i1997-00436-6},
	journal = {Europhysics Letters},
	month = {oct},
	number = {2},
	pages = {141},
	publisher = {},
	title = {Dissipative particle dynamics with energy conservation},
	url = {https://doi.org/10.1209/epl/i1997-00436-6},
	volume = {40},
	year = {1997}
}

@article{PEspanol_1997,
	author = {P. Español},
	doi = {10.1209/epl/i1997-00515-8},
	journal = {Europhysics Letters},
	month = {dec},
	number = {6},
	pages = {631},
	publisher = {},
	title = {Dissipative particle dynamics with energy conservation},
	url = {https://doi.org/10.1209/epl/i1997-00515-8},
	volume = {40},
	year = {1997}
}

@article{Caspers_2023,
    author = {Caspers, Juliana and Ditz, Nikolas and Krishna Kumar, Karthika and Ginot, Félix and Bechinger, Clemens and Fuchs, Matthias and Krüger, Matthias},
    title = {How are mobility and friction related in viscoelastic fluids?},
    journal = {The Journal of Chemical Physics},
    volume = {158},
    number = {2},
    pages = {024901},
    year = {2023},
    month = {01},
    issn = {0021-9606},
    doi = {10.1063/5.0129639},
    url = {https://doi.org/10.1063/5.0129639}
}

@article{Vaidya_2025,
  title = {Observation and control of nonmonotonic recoils in a viscoelastic fluid},
  author = {Vaidya, Salil S. and Muruga, Lokesh and Caspers, Juliana and Kr\"uger, Matthias and Bechinger, Clemens},
  journal = {Phys. Rev. Res.},
  volume = {7},
  issue = {3},
  pages = {033084},
  numpages = {6},
  year = {2025},
  month = {Jul},
  publisher = {American Physical Society},
  doi = {10.1103/74nx-3t1h},
  url = {https://link.aps.org/doi/10.1103/74nx-3t1h}
}

@article{CALDEIRA1983587,
	author = {A.O. Caldeira and A.J. Leggett},
	doi = {10.1016/0378-4371(83)90013-4},
	issn = {0378-4371},
	journal = {Physica A: Statistical Mechanics and its Applications},
	number = {3},
	pages = {587–616},
	title = {Path integral approach to quantum Brownian motion},
	url = {https://www.sciencedirect.com/science/article/pii/0378437183900134},
	volume = {121},
	year = {1983}
}

@article{Cao2023,
  author    = {Cao, Xin and Das, Debankur and Windbacher, Niklas and Ginot, F{\'e}lix and Kr{\"u}ger, Matthias and Bechinger, Clemens},
  title     = {Memory-induced Magnus effect},
  journal   = {Nature Physics},
  year      = {2023},
  volume    = {19},
  number    = {12},
  pages     = {1904--1909},
  doi       = {10.1038/s41567-023-02213-1},
  url       = {https://doi.org/10.1038/s41567-023-02213-1},
  issn      = {1745-2481}
}

@article{Guo2004EigenViscoelastic,
  author  = {Shao-hua, GUO},
  title   = {Eigen Theory of Viscoelastic Dynamics Based on the Kelvin--Voigt Model},
  journal = {Applied Mathematics and Mechanics (English Edition)},
  year    = {2004},
  volume  = {25},
  number  = {7},
  pages   = {792--798}
}

@article{Trcala2024,
  author    = {Trcala, M. and Suchomelov{\'a}, P. and Bo{\v{s}}ansk{\'y}, M. and Hoke{\v{s}}, F. and N{\v{e}}mec, I.},
  title     = {The generalized Kelvin chain-based model for an orthotropic viscoelastic material},
  journal   = {Mechanics of Time-Dependent Materials},
  year      = {2024},
  volume    = {28},
  number    = {3},
  pages     = {1639--1659},
  doi       = {10.1007/s11043-024-09678-4},
  url       = {https://doi.org/10.1007/s11043-024-09678-4},
  issn      = {1573-2738}
}

@article{PHANTHIEN20166359,
	author = {N. Phan-Thien and N. Mai-Duy and B.C. Khoo and D. Duong-Hong},
	doi = {10.1016/j.apm.2016.01.051},
	issn = {0307-904X},
	journal = {Applied Mathematical Modelling},
	number = {13},
	pages = {6359–6375},
	title = {Strongly overdamped Dissipative Particle Dynamics for fluid-solid systems},
	url = {https://www.sciencedirect.com/science/article/pii/S0307904X16300476},
	volume = {40},
	year = {2016}
}

@article{Tong2023,
  title = {Linear viscoelastic response of the vertex model with internal and external dissipation: Normal modes analysis},
  author = {Tong, Sijie and Sknepnek, Rastko and Ko\ifmmode \check{s}\else \v{s}\fi{}mrlj, Andrej},
  journal = {Phys. Rev. Res.},
  volume = {5},
  issue = {1},
  pages = {013143},
  numpages = {20},
  year = {2023},
  month = {Feb},
  publisher = {American Physical Society},
  doi = {10.1103/PhysRevResearch.5.013143},
  url = {https://link.aps.org/doi/10.1103/PhysRevResearch.5.013143}
}

@article{SerraAguila2019,
  author  = {Serra-Aguila, A. and Puigoriol-Forcada, J. M. and Reyes, G. and Menacho, J.},
  title   = {Viscoelastic models revisited: characteristics and interconversion formulas for generalized Kelvin--Voigt and Maxwell models},
  journal = {Acta Mechanica Sinica},
  year    = {2019},
  volume  = {35},
  number  = {6},
  pages   = {1191--1209},
  doi     = {10.1007/s10409-019-00895-6},
  url     = {https://doi.org/10.1007/s10409-019-00895-6},
  issn    = {1614-3116}
}

@article{godreche2018characterising,
  title={Characterising the nonequilibrium stationary states of Ornstein--Uhlenbeck processes},
  author={Godr{\`e}che, Claude and Luck, Jean-Marc},
  journal={Journal of Physics A: Mathematical and Theoretical},
  volume={52},
  number={3},
  pages={035002},
  year={2018},
  publisher={IOP Publishing},
  doi={10.1088/1751-8121/aaf190},
  url={https://iopscience.iop.org/article/10.1088/1751-8121/aaf190/pdf}
}

@article{Kubo1966FDT,
	author = {R Kubo},
	doi = {10.1088/0034-4885/29/1/306},
	journal = {Reports on Progress in Physics},
	month = {jan},
	number = {1},
	pages = {255},
	publisher = {},
	title = {The fluctuation-dissipation theorem},
	url = {https://doi.org/10.1088/0034-4885/29/1/306},
	volume = {29},
	year = {1966}
}

@article{Tsvi_Tlusty2018Protein,
	author = {Sandipan Dutta and Jean-Pierre Eckmann and Albert Libchaber and Tsvi Tlusty},
	doi = {10.1073/pnas.1716215115},
	journal = {Proceedings of the National Academy of Sciences},
	number = {20},
	pages = {E4559–E4568},
	title = {Green function of correlated genes in a minimal mechanical model of protein evolution},
	url = {https://www.pnas.org/doi/abs/10.1073/pnas.1716215115},
	volume = {115},
	year = {2018}
}

@article{RevModPhys.91.031001,
  title = {Colloquium: Proteins: The physics of amorphous evolving matter},
  author = {Eckmann, Jean-Pierre and Rougemont, Jacques and Tlusty, Tsvi},
  journal = {Rev. Mod. Phys.},
  volume = {91},
  issue = {3},
  pages = {031001},
  numpages = {28},
  year = {2019},
  month = {Jul},
  publisher = {American Physical Society},
  doi = {10.1103/RevModPhys.91.031001},
  url = {https://link.aps.org/doi/10.1103/RevModPhys.91.031001}
}

@article{Brandenbourger2019,
  author    = {Brandenbourger, Martin and Locsin, Xander and Lerner, Edan and Coulais, Corentin},
  title     = {Non-reciprocal robotic metamaterials},
  journal   = {Nature Communications},
  year      = {2019},
  volume    = {10},
  number    = {1},
  pages     = {4608},
  doi       = {10.1038/s41467-019-12599-3},
  url       = {https://doi.org/10.1038/s41467-019-12599-3},
  issn      = {2041-1723}
}

@article{Scheibner2020OddElasticity,
  author  = {Scheibner, Colin and Souslov, Anton and Banerjee, Debarghya and Sur{\'o}wka, Piotr and Irvine, William T. M. and Vitelli, Vincenzo},
  title   = {Odd elasticity},
  journal = {Nature Physics},
  year    = {2020},
  volume  = {16},
  number  = {4},
  pages   = {475--480},
  doi     = {10.1038/s41567-020-0795-y},
  url     = {https://doi.org/10.1038/s41567-020-0795-y},
  issn    = {1745-2481}
}

@article{Kailasham2026DPD,
  title = {Dissipative response of driven bead-spring-dashpot chains},
  author = {Kailasham, R.},
  journal = {Phys. Rev. E},
  volume = {113},
  issue = {6},
  pages = {065408},
  numpages = {18},
  year = {2026},
  month = {Jun},
  publisher = {American Physical Society},
  doi = {10.1103/pgt9-xtfl},
  url = {https://link.aps.org/doi/10.1103/pgt9-xtfl}
}

@book{goldstein_mechanics,
  author = {Goldstein, Herbert and Poole, Charles and Safko, John},
  year = {2002},
  title = {Classical Mechanics},
  edition = {3rd},
  publisher = {Addison Wesley},
  address = {Boston}
}

@book{kittel2004introduction,
  title={Introduction to Solid State Physics},
  author={Kittel, C.},
  isbn={9780471415268},
  lccn={2004042250},
  url={https://books.google.de/books?id=kym4QgAACAAJ},
  year={2004},
  publisher={Wiley}
}

@book{risken1996fokker,
  title={The Fokker-Planck Equation},
  author={Risken, Hannes},
  year={1996},
  publisher={Springer}
}

@book{landau1987fluid,
  title={Fluid Mechanics: Volume 6},
edition = {2nd},
  author={Landau, Lev Davidovich and Lifshitz, Evgeny Mikhailovich},
  year={1987},
  publisher={Elsevier}
}

@book{hansen2013theory,
  title={Theory of Simple Liquids: with Applications to Soft Matter},
  author={Hansen, Jean-Pierre and McDonald, Ian Ranald},
  year={2013},
  publisher={Academic Press}
}

@book{doi1988theory,
  title={The Theory of Polymer Dynamics},
  author={Doi, Masao and Edwards, Samuel Frederick},
  volume={73},
  year={1988},
  publisher={Oxford University Press}
}

@book{axler2024linear,
  title={Linear Algebra Done Right},
  author={Axler, Sheldon},
  year={2024},
  publisher={Springer}
}

@book{arfken2011mathematical,
  title={Mathematical Methods for Physicists: A Comprehensive Guide},
  author={Arfken, George B and Weber, Hans J and Harris, Frank E},
  year={2011},
  publisher={Academic press}
}

@misc{weissteinCongruence,
  author = {Weisstein, Eric W.},
  title = {Congruence Transformation},
  note = {From MathWorld--A Wolfram Resource},
  howpublished = {\url{https://mathworld.wolfram.com/CongruenceTransformation.html}},
}

@article{Taylor1923,
    author = {Taylor, Geoffrey Ingram},
    title = {VIII. Stability of a viscous liquid contained between two rotating cylinders},
    journal = {Philosophical Transactions of the Royal Society of London, Series A: Containing Papers of a Mathematical or Physical Character},
    volume = {223},
    number = {605-615},
    pages = {289-343},
    year = {1923},
    month = {01},
    issn = {0264-3952},
    doi = {10.1098/rsta.1923.0008},
    url = {https://doi.org/10.1098/rsta.1923.0008}
}

@misc{Niloy2026,
  author = {Roy, Niloyendu and Saha, Rupayan and  Das, Debankur and Kr{\"u}ger, Matthias and Bechinger, Clemens },
  title  = {Geometry-Controlled Relaxation Spectra in a Viscoelastic Fluid},
  year   = {2026},
  note   = {In preparation}
}

\end{document}